# Linking Publications to Funding at Project Level:
*A curated dataset of publications reported by FP7 projects*

Alexis-Michel Mugabushaka[1]

[1] *alexis-michel.mugabushaka@ec.europa.eu*
European Research Council Executive Agency (ERCEA), Covent Garden Building, Place Charles Rogier 16
1210 Bruxelles (Belgium)

**Abstract**

Datasets explicitly linking publications to funding at project level are the basis of evaluative bibliometric analysis of funding programmes. Analysis of the impact of the EU funding programmes has been often frustrated by the lack of data on publications to which the funding has contributed. Here we present a dataset[2] of scholarly publications reported by the projects funded by the European Union under the 7th Framework Programme. The dataset was created by first consolidating data from different reporting channels and validating the records by systematically matching them to external authoritative sources and assigning them external identifiers.

The initial dataset had 305k records linked to one or more projects out of which 69% had a digital object identify (doi). Through the data quality assurance, we validate 93% of the initial records (283k) and assign a doi to 90 % of them of them (245k). The resulting dataset has 245k unique dois (linked to one or more projects). It is, to our knowledge, the first comprehensive and curated dataset of scholarly outputs of the Framework Programme as reported by the grant holders.

The dataset could only be created thanks to significant improvements and investments made in the reporting systems used by EU funded projects.

**Introduction**

Bibliometric analysis has been an integral part of evaluation of research funding programmes for decades.

An example of one of the earliest studies is reported in Francis Narin seminal book "*evaluative bibliometrics*" (Narin, 1976). He analyzed the relationship between NIH funding and the number publications for over 200 institutions receiving about 90% of NIH funding between 1965 and 1972). He found a strong linear relationship between the number of publications and funds awarded. In his view, the finding - indicating that large institutions and moderately sized institutions tend to behave similarly in terms of "*utilizing funds to produce research output*"- contradicted an intuitive expectation based on the "*the economy of scale*". (i.e. with larger organizations producing marginally less number of publications).  However, it can even be argued that the evaluation of funding programmes, from peer review to differences in performance between different programmes played a key role in the development of Scientometrics as a research field. Indeed, as detailed in the analysis of Megan Jendrysik (Jendrysik, M. 2020), the US National Institutes of Health (NIH),  and National Science Foundation (NSF) funded a proposal of Eugene Garfield in early 60s to create the maps of science and laid the foundation to the Citation Index. The initial funding was focused on a citation index for research in Genetics but the resulting Science Citation Index (SCI) was soon used by both NIH and NSF to report the effectiveness of their funding strategies to the US Congress and the public.

---

[1] The views expressed in this paper are the author's. They do not necessarily reflect the views or official positions of the European Commission, the European Research Council Executive Agency or the ERC Scientific Council.

[2] Dataset available EU open data portal: https://data.europa.eu/data/datasets/cordisfp7projects

The availability of funding acknowledgments in commercial bibliometric databases (Web of Science since 2008, Scopus since 2017 and Dimensions since 2018) and from other sources such as Medline, EuroPMC, OpenAire etc... has triggered interest in this type of analysis linking funding to publications. Recent examples making use of those data include the weight of the NIH in funding cardiovascular disease research (Lybarova et. al. 2009) or the impact of the nanotechnology funding boom between 2000 and 2010 (Shapira & Wang 2010).

Studies using publications at the level of funding agencies or funding programmes can generate useful information on the results of funding. They are however of limited value in getting a deep understanding of the impact of funding (and in generating actionable insights which can be used for example to adjust/change funding conditions). Usually all major funding agencies have multiple funding mechanisms, which differ, not only in their stated objectives but also in criteria used to select among competing proposals and in the terms and conditions of funding. By missing those differences, bibliometric analysis at the level of the funding agency or programmes cannot inform for example on the extent to which the funding has contributed to its stated objectives. They also cannot untangle factors explaining differences in publication patterns within the same funding scheme. Such nuanced analyses are only possible with data linking grants and publications. This has been long recognized by bibliometricians but as Kevin Boyack remarks, was not done routinely because "*data explicitly linking articles with the grants from which they were funded (were) lacking*" (Boyack 2009).

This has been particularly the case for the European Union funding programmes for many years. Several projects funded in the context of its evaluation, produced interesting insights on the participants' profiles and motives as well as on how they perceive the effects and impact of funding. However, those analysis have been often frustrated by the lack of data on results of projects and in particular of publications to which the funding has contributed. As one project – complaining about the fact that the available data did not allow them to go beyond the description of the projects – puts it. "*While we were able to categorize the health-related projects into themes on the basis of their titles, we could say little else as there was almost no information on results or conclusions of projects, with records instead describing the work that was proposed*". (Ernst et al. 2010, p. 135).

Starting from the Seventh Framework Programme (2007-2014), significant efforts have been made in both the reporting of project results to the European Commission services and external infrastructures such as OpenAire, which harvest open science systems to create datasets of publications from EU funded projects. Here we present a curated dataset of publications reported by grant holders from projects funded by the European Union (under the Seventh Research Framework programme) which builds on those efforts.

We start in the next section by discussing various possibilities to link publications to funding at project level. In the subsequent section we present the dataflow and quality assurance process used in creating the dataset. The concluding section reflects on the limitations of both our approach and its implication for the quality of the dataset.

## 2. Grants - publications links: various mechanisms

It is important to note that reporting is only one of several possible mechanisms through which publications are linked to grants. Other channels are acknowledgments, institutional repositories and through principal investigators (see Figure 1). Each of those mechanisms has its advantages and disadvantages that we discuss below.

- The channel, which has received much attention, is the <u>*acknowledgment.*</u> It is potentially less burdensome as the grant holders have only to indicate the funding once. It also has some limitations. Some studies reveal some open data quality issues – in terms of completeness (at project level) and recording in consistent manner- as well as different acknowledgments practice across research areas and research systems (Sirtes, D. (2013), Álvarez-Bornstein et al. (2017)) or varying acknowledgment requirements by funding agencies. There have also been some concerns on the risk of "*strategic acknowledgment*". As J. Rigby (2011) puts it, in the acknowledgment, authors may bring their own agendas. "*Perceived low-status funding bodies may be left off the list of acknowledgements (…) conversely, high-status funding bodies might find that they were listed on more of the papers of a group of authors than were genuinely produced with their help simply to enhance the reputation of the authors*".

- *Reporting* is the second channel through which publications can be linked to grants. Most – if not all – funding agencies require grant holders to report publications to which the funding has contributed. The link is straightforward as the grant holders certify it. The disadvantage of this method is that the efforts needed to "clean" the reported publications can be substantial if they are not properly recorded. On the other hand, the risk of strategic behaviours ("strategic reporting") can also not be ruled out: grant holders selectively reporting some items (e.g. including those to which the funding has not contributed) or leaving out those which may raise questions in terms of compliance with funding terms (e.g. those not respecting a funder's open access mandate). Another issue is that reporting often captures only the publications produced before the end of the grant.

- A third linkage channel are the repositories which were set to support open access/open science. Those "*open science repositories*" whether institutional or subject specific (like EuropePMC often include the source of funding. One infrastructure which systematically harvests and curates data from repositories is OpenAire (Manghi et al. (2010). OpenAire is pan-European research information system, which started in 2006 as an initiative to network Open Access repositories (funded by the European Commission as project Driver), and has evolved into a full-fledged infrastructure to support the European Union Open science agenda.

- Another way, which can be considered, is to link publications to funding through the *Principal Investigator*, by considering the scholarly outputs during grant time to be a result of funding. This would be the simplest way to make the link (one would only need to identify the PI in bibliographic databases). It is however only meaningful for funding programmes which are person focused and are the only (or at least the major) source of funding for the person such as the Howard Hughes Medical Institute (HHMI).

**Table 1: Commonly used Mechanisms for linking publications to grants (at project level).**

| | Advantages | | Data Quality | Limitations |
|---|---|---|---|---|
| Acknowledgment | Retrieve funding information from acknowledgment sections of publications | Data available in commercial databases and other sources | Link to specific grants/projects not always made | Risk of strategic acknowlegment |
| Reporting | Grant holders provididing list of publications to which the funding has contributed | Link made by grant holders | Timing : post grant publications not always recorded | Risk of strategic reporting |
| Open Science Repositories | Records in open science repositories linked to source of funding | Available in institutional or subject specific repositories | not assessed | Coverageor completness not clear |
| Through Principal Investigators (PI) | Using entire scholarly production of funded PI as enabled by the funding | Data available in commercial databases and other sources | Can be relatively high | Valid only for PI-focused funding |

In considering those publication-to-grant linkage mechanisms, there are two caveats to keep in mind.

- First, although they are presented as distinct in reality they might not be. For example, while Web of Science started between publications with funding, using only acknowledgments, it has recently moved to include also funding information from reporting in particular ResearchFish and Medline (Liu et al (2020). This almost certainly the case for other commercial databases although they are less transparent about the origin of their funding data. Systematic comparisons between those sources could inform on the extent to which they overlap or complement each other. An important implication here is that caution is needed if funding information from those commercial databases is used in comparing funding agencies/programmes. The basis may simply not be the same if for one the acknowledgments in publications is used and for others the publications reported by the grant holders are included. For example, any study based on the acknowledgments from Web of Science Clarivate will almost certainly have a higher number of publications for UK research councils which use ResearchFish and less for other funding agencies.
- The second caveat concerns the assumption underlying the linking of publications to grants: namely the idea that a publication is somehow a "*product*" of the grant. This is clearly an oversimplification, which can be challenged as a "*project fallacy assumption*" (i.e the idea that a project funding results in well-defined and attributable -uniquely- sets of publications/results). In reality, publications/results are often products of longer running research undertakings to which multiple sources of funding contribute simultaneously or sequentially.

## 3. Publications reported by EC funded projects in FP7 : data sources and data processing flow

Grant holders from EU funded projects are required to report, among other, scientific publications to which the funding has contributed. For the FP7 programme, the publications are reported with every reporting period roughly twice within the life cycle of a project (the reporting cycle varies across programmes and funding schemes).

Grant holders report publications through an online system (called SESAM). Since the beginning, all parts of the Programme used this system with some exceptions:

- The European Research Council (ERC) which has a slightly different reporting cycle and started using it in 2011.
- The Cooperation Programme ICT collected the publications through a series of annual surveys of project coordinators (Jacob et al. 2016).
- Some parts of the Framework Programmes which operates for example in "indirect funding mode" (i.e. allocated grants which they subsequently use to fund specific projects) which did not report at the project level in the reporting system. This includes some co-funded programmes such as Joint Technology Initiatives (JTIs).

Accordingly, we have three major sources of publications data: (1) the common reporting system, the (2) ERC data recorded before it started using the common reporting system and (3) data from the ICT Cooperation programmes.

To create a consolidated dataset, the data from those three sources[3] were imported in a database (PostgreSQL) and underwent a data quality processing consisting of three major steps which are briefly described below.

---

[3] We used data available as of 21st August 2021

**Figure 1. Data sources and Data Flow**

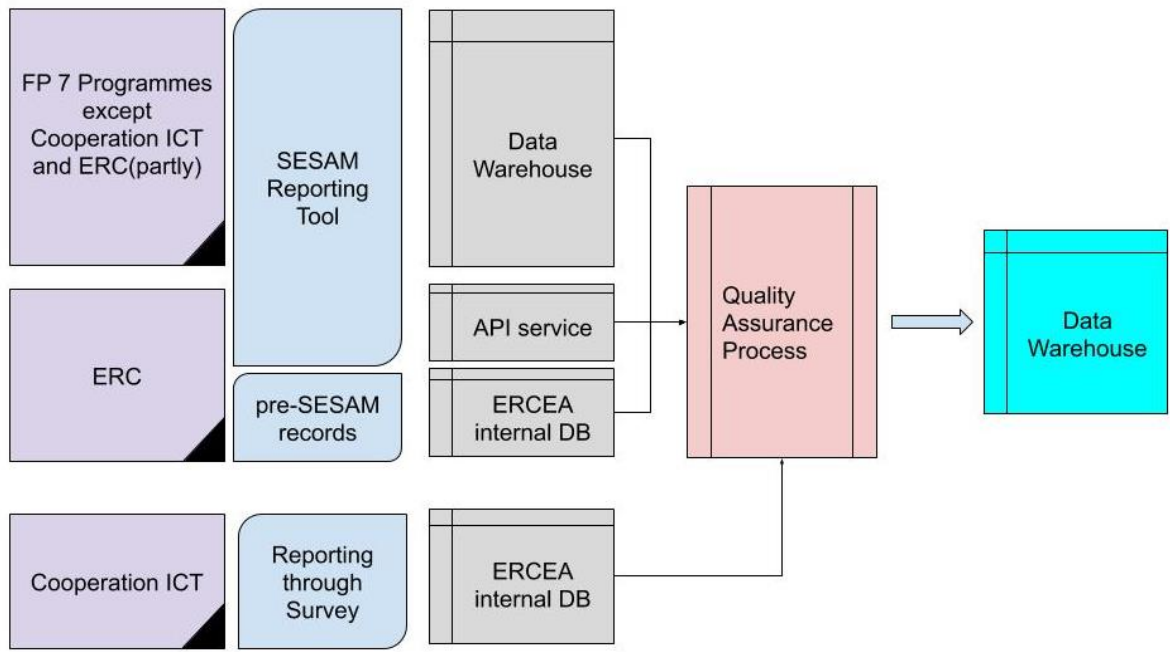

*3.1 Validation of entries*

Records reported by grant holders as publications encompass all types of publications including working papers, which are subsequently published as articles, project promotional brochures etc... They may also include records, which have been submitted but not (yet) accepted at the time of reporting.

The validation step consists in detecting genuine scholarly publications. The approach chosen was to validate all reported records against external authoritative sources. This is done by means of matching those records to selected sources and attributing them a validated identifier from those sources.

In this process, the most important validation "authority" is Crossref. Crossref is a registration agency whose membership includes virtually all scholarly publishers and it issues digital object identifiers (doi) as persistent identifiers for scholarly outputs (Hendricks, G. et al 2020). Although it is not the agency registering digital object identifiers (dois), for the publications under consideration here it is by far the most important one.

For the FP7 programme, the reporting system foresees an automatic retrieval of publication metadata from Crossref. Grant holders can enter a doi and the system retrieves metadata from Crossref. This process was applied only to certain types of publications (journal articles) and in some cases, it seems not to have been used at all. A cursory inspection of records showed that some entries without a doi in the system have in fact a valid doi in crossref.

The first task in the validation step consists in verifying the recorded digital object identifiers (dois). For this we use a regular expression recommended by Crossref to filter out the entries, which are not well-formed dois. … In addition, we also verify if those well well-formed dois can be matched to an external source. This step was introduced after we realized that even among records which pass the filter through the regular expression, some turned out to be non-valid dois. As an external source, we use a local copy of Microsoft academic Graph[4] (Sinha et al. 2015 and Wang, K 2019).

*3.2 Finding missing digital object identifiers*

Some of the records without a recorded doi are in fact genuine scholarly publications published in well-established journals and conferences. The next step is to find missing dois and that where the bulk of the work in the quality assurance process goes.

Our initial approach was to use *Crossref api* to match records to their dois. Crossref offers a service which accepts as an input an unstructured string representing a publication and returns its doi or more precisely candidate records with doi as well as a similarity measure between the input and a candidate match. It has been shown to perform well on different benchmarks.

In our case we used as an input a string combining authors, titles of the paper, journal or conference as well as publication year and retrieved up to top three candidate matches (with respect to similarity measure provided by Crossref. By looking in the returned records we realized that the service was not performing as well as we expected.

- One problem was our input strings which were not always complete (for example missing author names or listing only one author) or well formed (some had paper titles in the field for journals/conferences).
- Another problem was caused by the services when there were records with almost identical titles. This is the case for example for reviews of books (which often repeat the reviewed book in the titles), corrections to articles (errata) or the development in some journals to invite for comments/post-publication reviews on selected articles (we observed this often in Geosciences). In that case it was difficult to match the records in our system and the records in crossref based on ranking by similarity.

We then decided to also use for this step a local copy of the Microsoft academic graph. We indexed (in Elasticsearch) about 100 Mio records (corresponding to publications from the period after 2007) and then used string similarity search to match them to MAG records and through MAG id to their dois.

**Figure 2. Data Quality Assurance Process**

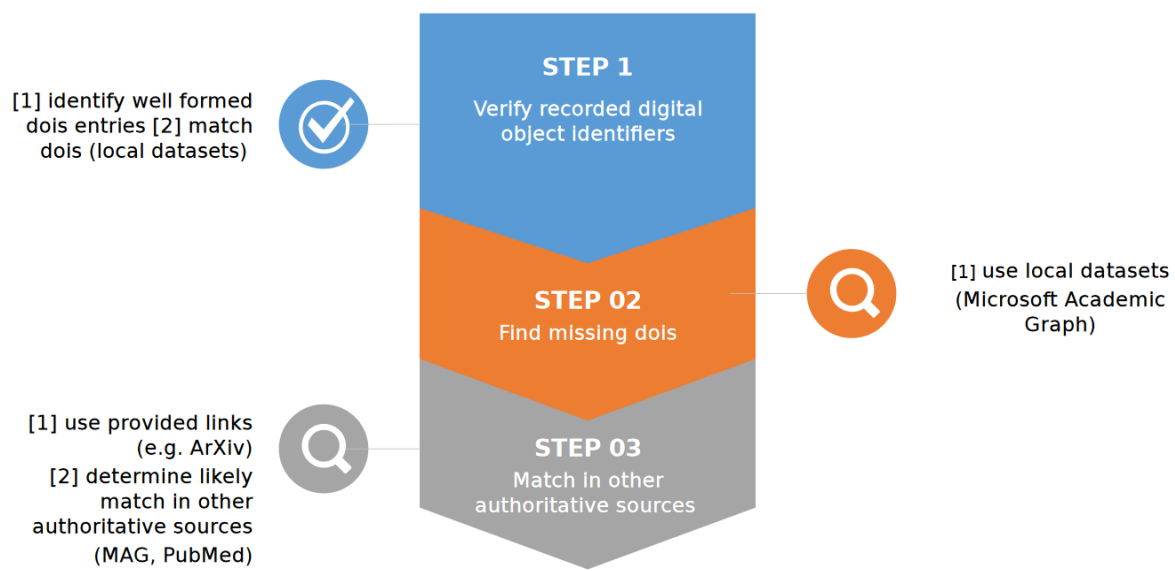

---

[4] Version extracted on 15th February 2021

*3.3 Validating against other authoritative sources*

In the last step, records for which a doi could not be found were matched to other authoritative sources and assigned their identifiers. In this process, a cascading process is applied. A publication is flagged as validated if it has been matched (in cascading order) to a record crossref via doi (first two steps), Microsoft Academic Graph (via MAG identifier), OpenAire or Arxiv. Other sources are also considered – though not systematically – such as SCOPUS, Web of Science, PubMed but also institutional repositories or Amazon (for books).

The three steps in the QA process allows not only to validate reported records (ultimately confirming that they are genuine publications or not). They also serve the purpose of the deduplication of records. A record is considered a duplicate if the publication is reported more than once for the same project. In this case the last record (in terms of entry date) is retained in the dataset and others are flagged as duplicates. Finally, the validation by matching to external sources allows also to retrieve further data on the basis of which further analysis can be made.

**4. Resulting Dataset**

As described earlier, the data quality assurance process started with three datasets: publications reported through SESAM, publications from projects of the ICT cooperation programme gathered through annual surveys and publications from ERC funded projects which were reported before ERC Grant holders could also use the SESAM reporting tool (ERC "pre-sesam publications"). The three datasets had respectively about 283k, 18k and 4k publications.

The table 2 shows the results of the data quality assurance process. In the first dataset about 68% of the records had originally dois. Through the data quality assurance process this share raised to about 90%. For the IC dataset the share remained fairly the same: a fact which can be explained the extensive data quality assurance undertaken before the release of the data (Jakob et al. 2016).

Putting all the datasets together: from the initial 305,549 records, the resulting dataset includes over 283k validated records, 278k of which were validated by matching them to a doi. The unique number of dois is about 245k.

**Table 2: Results of Data Quality Assurance Process**

|  | all datasets | ERC pre-SESAM dataset | ICT dataset | SESAM dataset |
|---|---:|---:|---:|---:|
| all reported records | 305,549 | 4,200 | 18,158 | 283,191 |
| records reported with doi (validated as such) | 209,513 | 2,294 | 16,466 | 190,753 |
| records validated through QA | 283,095 | 2,849 | 18,033 | 262,213 |
| records validated through QA with doi | 272,923 | 2,294 | 17,030 | 253,599 |
| records validated through QA - unique records | 255,075 | 2,756 | 18,031 | 235,179 |
| records validated through QA - unique doi | 245,337 | 2,220 | 17,028 | 226,977 |
| % records reported with doi (validated as such) | 68.6 | 54.6 | 90.7 | 67.4 |
| % records validated through QA | 92.7 | 67.8 | 99.3 | 92.6 |
| % records validated through QA with doi | 89.3 | 54.6 | 93.8 | 89.6 |

## 5. Concluding remarks

We present a dataset of publications reported by projects funded by the European Union under the 7th Framework (2007-2013). It consolidates data from different reporting channels and systematically matches them to authoritative external data sources. The consolidated dataset includes about 245k publications with unique dois.

The consolidated dataset was the result of a long and labor-intensive data quality assurance process, but it could only be produced because of the significant improvements made in the reporting tools by EU funded projects. Indeed, there are no systematic publications records from previous frameworks programmes and only changes made in F7 reporting mechanisms make this effort possible. We note that the publications reported by the projects funded under the H2020 programme (2014-2020) are also accessible through the EU open data portal. Some of the data quality approaches tested here can also be used for this dataset.

This dataset has also a number of limitations:

- As mentioned in section 2, using the reporting channel leads to an underestimate of publications records as some appears long after the project has "administratively" closed. This is partly mitigated by the fact that some projects reported submitted papers or pre-prints, which could be matched to the published records, but discrepancies, will remain. In case of the FP7, we also note that projects for some parts of the programs were using different reporting channels. For this reason, this dataset cannot be considered a comprehensive and definitive dataset of project publications but rather a more or less complete records of those reported during the lifetime of the project.
- As of August 2021, there about 11 projects still running and whose (publication) results are not included in the dataset. Those should be integrated as soon as their records are available.
- Finally, a dataset of this size cannot be free of mistakes. Especially the QA process which was largely automated is likely to introduce errors of its own. Here we hope to rely on the users who will report the mistakes found so that they can be subsequently corrected.

To our knowledge, this is the first comprehensive and curated dataset of reported publications by EU funding. We hope that will be useful to different stakeholders with an interest in the results of European Union funding and beyond.

## Acknowledgment

The results reported here would not have been possible without the support of my colleagues Daniel Szmytkowski and Lucas Holgado Guillen from the European Commission Directorate General for Research and Innovation (DG RTD). The work benefited greatly from their expertise and dedication. They also saw and commented on this paper in its draft form but were not asked explicitly asked to endorse its approach or interpretations. Any shortcomings of the paper are the authors' alone.

## 6. References

Álvarez-Bornstein, B., Morillo, F., & Bordons, M. (2017). Funding acknowledgments in the Web of Science: completeness and accuracy of collected data. Scientometrics, 112(3), 1793-1812
https://doi.org/10.1007/s11192-017-2453-4

Boyack, K. W. (2009, July). Linking grants to articles: Characterization of NIH grant information indexed in Medline. In 12th International Conference of the International Society for Scientometrics and Informetrics (Vol. 2009, pp. 730-41)

Ernst, K., Irwin, R., Galsworthy, M., McKee, M., Charlesworth, K., & Wismar, M. (2010). Difficulties of tracing health research funded by the European Union. Journal of health services research & policy, 15(3), 133-136. https://doi.org/10.1258/jhsrp.2010.009115

Hendricks, G., Tkaczyk, D., Lin, J., & Feeney, P. (2020). Crossref: The sustainable source of community-owned scholarly metadata. Quantitative Science Studies, 1(1), 414-427. https://doi.org/10.1162/qss_a_00022

Jacob, J., Sanditov, B., Smirnov, E., Wintjes, R., Surpatean, A., Notten, A., & Sasso, S. (2016). Analysis of publications and patents of ICT research in FP7. Brussels: European Commission https://op.europa.eu/en/publication-detail/-/publication/2f800262-ea80-11e5-a2a7-01aa75ed71a1

Jendrysik, Meghan A. 2020. The Role of the NIH in the Developmentand Promotion of Citation Indexing and Analysis as ScientometricTools. Master's thesis, Harvard Extension School. https://nrs.harvard.edu/URN-3:HUL.INSTREPOS:3736488

Liu, W., Tang, L. & Hu, G. Funding information in Web of Science: an updated overview. Scientometrics 122, 1509–1524 (2020). https://doi.org/10.1007/s11192-020-03362-3

Lyubarova, R., Itagaki, B. K., & Itagaki, M. W. (2009). The impact of National Institutes of Health funding on US cardiovascular disease research. PLoS One, 4(7), e6425. https://doi.org/10.1371/journal.pone.0006425

Manghi, P., Manola, N., Horstmann, W., & Peters, D. (2010). An infrastructure for managing EC funded research output-The OpenAIRE Project. The Grey Journal (TGJ): An International Journal on Grey Literature, 6(1).

Narin, F. (1976). Evaluative bibliometrics: The use of publication and citation analysis in the evaluation of scientific activity (pp. 334-337). Cherry Hill, NJ: Computer Horizons. http://www.worldcat.org/oclc/3124746

Rigby, J. (2011). Systematic grant and funding body acknowledgement data for publications: new dimensions and new controversies for research policy and evaluation. Research Evaluation, 20(5), 365-375. https://doi.org/10.3152/095820211X13164389670392

Sinha, A., Shen, Z., Song, Y., Ma, H., Eide, D., Hsu, B. J., & Wang, K. (2015, May). An overview of microsoft academic service (mas) and applications. In Proceedings of the 24th international conference on world wide web (pp. 243-246). https://doi.org/10.1145/2740908.2742839

Sirtes, D. (2013). Funding acknowledgements for the German research foundation (Dfg): the dirty data of the Web of Science database and how to clean it up. In Pro Int Conf Sci Inf (pp. 784-95) https://doi.org/10.1007/s11192-017-2453-4

Wang, J., & Shapira, P. (2011). Funding acknowledgement analysis: an enhanced tool to investigate research sponsorship impacts: the case of nanotechnology. Scientometrics, 87(3), 563-586. https://doi.org/10.1007/s11192-011-0362-5

Wang, K., Shen, Z., Huang, C. Y., Wu, C. H., Eide, D., Dong, Y., ... & Rogahn, R. (2019). A review of Microsoft academic services for science of science studies. Frontiers in Big Data, 2, 45. https://doi.org/10.3389/fdata.2019.00045